\pdfoutput=1

\documentclass[11pt,twoside,a4paper,cmspaper,final,collab]{cms-tdr}
\def\svnVersion{403859}\def\cmsCernNoTag{CERN-EP/2016-269}\def\cmsCernDate{\today}\def\cmsMessage{Published in Physics Letters B as \href{http://dx.doi.org/10.1016/j.physletb.2017.04.005}{\doi{10.1016/j.physletb.2017.04.005.}}}

\begin{document}\cmsNoteHeader{SUS-15-012}

\hyphenation{had-ron-i-za-tion}
\hyphenation{cal-or-i-me-ter}
\hyphenation{de-vices}
\RCS$Revision: 397198 $
\RCS$HeadURL: svn+ssh://svn.cern.ch/reps/tdr2/papers/SUS-15-012/trunk/SUS-15-012.tex $
\RCS$Id: SUS-15-012.tex 397198 2017-04-03 00:47:57Z areinsvo $
\newlength\cmsFigWidth
\ifthenelse{\boolean{cms@external}}{\setlength\cmsFigWidth{0.48\textwidth}}{\setlength\cmsFigWidth{0.67\textwidth}}
\ifthenelse{\boolean{cms@external}}{\providecommand{\cmsLeft}{top\xspace}}{\providecommand{\cmsLeft}{left\xspace}}
\ifthenelse{\boolean{cms@external}}{\providecommand{\cmsRight}{bottom\xspace}}{\providecommand{\cmsRight}{right\xspace}}
\cmsNoteHeader{SUS-15-012}
\title{Search for supersymmetry in events with photons and missing transverse  energy in pp collisions at 13\TeV}

\date{\today}

\abstract{
   The results of a search for new physics in final states with photons and
   missing transverse energy are reported. The study is based on a sample of
   proton-proton collisions collected at a center-of-mass energy of
   13\TeV with the CMS detector in 2015, corresponding to an integrated
   luminosity of 2.3\fbinv. Final states with two photons
   and significant missing transverse energy are used to search
   for supersymmetric particles in models of supersymmetry (SUSY) with general
   gauge-mediated (GGM) supersymmetry breaking. No
   excess is observed with respect to the standard model expectation, and the
   results are used to set limits on gluino pair production and squark pair
   production in the GGM SUSY framework. Gluino masses below 1.65\TeV and
   squark masses below 1.37\TeV are excluded at a 95\% confidence level.
}

\hypersetup{%
pdfauthor={CMS Collaboration},%
pdftitle={Search for supersymmetry in events with photons and missing  transverse energy in pp collisions at 13 TeV},%
pdfsubject={CMS},%
pdfkeywords={CMS, physics, supersymmetry, photons and missing transverse energy}}

\maketitle

\section{Introduction}
\label{sec:Introduction}

Final states in proton-proton collisions containing photons with high
transverse momentum \pt and significant missing transverse energy \ETmiss
emerge naturally from a variety of new-physics scenarios, particularly in
models of supersymmetry (SUSY) broken via gauge mediation that require a
stable, weakly interacting lightest supersymmetric particle
(LSP)~\cite{Nappi:1982bc, Dine:1982cd, Alvarez:1982de, Dine:1993yw, Dine:1994vc,
Dine:1995ag}. The \ETmiss in an event, defined as the magnitude of the vector
sum of the transverse momenta of all visible particles, is a consequence of
undetected particles such as neutrinos or LSPs. Models with general gauge
mediation (GGM)~\cite{Dimopoulos:1996fg, Martin:1997hi, Poppitz:1997gh, Meade:2008wd, 
Buican:2008ws, Abel:2009ve, Carpenter:2009jk, Dumitrescu:2010ef} can have a wide
range of features, but typically entail a nearly massless gravitino LSP,
\PXXSG, and a next-to-lightest supersymmetric particle (NLSP) often taken to be
a neutralino \PSGczDo. Photons in the final state arise when the neutralino 
decays to a gravitino and a photon, $\PSGczDo \to \PXXSG\PGg$.

\begin{figure}[htb]
    \centering
    \includegraphics[width=0.45\textwidth]{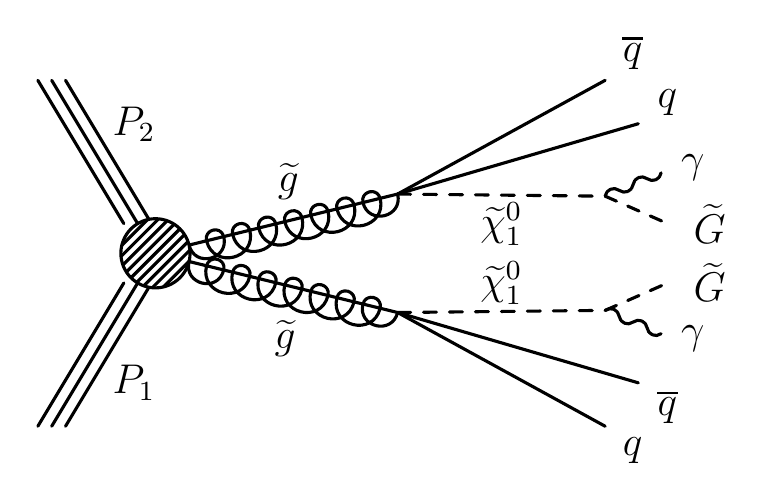}
    \includegraphics[width=0.45\textwidth]{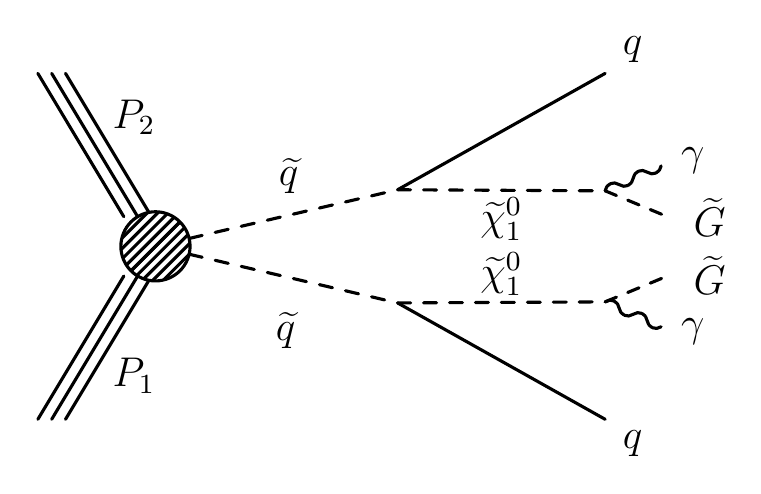}
    \caption{Diagrams showing the production of signal events in the collision
        of two protons with four momenta ${P}_{1}$ and ${P}_{2}$. In gluino
        \PSg~pair production in the T5gg simplified model (\cmsLeft), the gluino
        decays to an antiquark \cPaq, quark \cPq, and neutralino \PSGczDo. In
        squark \PSq~pair production in the T6gg simplified model (\cmsRight), the
        squark decays to a quark and a neutralino. In both cases, the
        neutralino subsequently decays to a photon \PGg and a gravitino \PXXSG.
        In the second diagram, we do not distinguish between squarks and
        antisquarks.}
    \label{fig:gluinoSquarkDecay}
\end{figure}

In this Letter we present a search for GGM SUSY in final states involving two
photons and significant \ETmiss. The data sample, corresponding to an
integrated luminosity of 2.3\fbinv of proton-proton collisions at $\sqrt{s} = 
13\TeV$, was collected with the CMS detector at the CERN LHC in 2015. The
increased center-of-mass energy substantially improves the sensitivity of the
analysis compared to searches performed at the LHC in Run 1 at
$\sqrt{s} = 8\TeV$~\cite{Aad:2015hea, Khachatryan:2015exa}. A similar analysis
was performed by the ATLAS Collaboration at $\sqrt{s} = 13\TeV$~\cite{ATLAS:2016aa}.
For the interpretation of the results we use the T5gg and T6gg simplified
models~\cite{expPaper}. The T5gg (T6gg) simplified model assumes gluino
\PSg~(squark \PSq) pair production, with subsequent decays as shown in
Fig.~\ref{fig:gluinoSquarkDecay}. The branching fraction of the NLSP neutralino
to decay to a gravitino and a photon, $\PSGczDo \to \PXXSG\PGg$,
resulting in characteristic events with two photons and large \ETmiss, is
assumed to be unity. In more general GGM SUSY models, a bino-like neutralino
could also decay to a gravitino and a Z boson, $\PSGczDo \to \PXXSG\cPZ$.

Events with two photons and \ETmiss can also arise from several standard model
(SM) processes, including direct diphoton production with initial-state
radiation and multijet events (possibly with associated photon production).
These processes lack intrinsic \ETmiss but can emulate the signal if the
hadronic activity in the event is mismeasured. In the latter case, photons may
be reconstructed in the event as a result of the misidentification of
electromagnetically rich jets. A smaller background comes from events with
intrinsic \ETmiss, principally $\PW\PGg$ and W+jet production, where an electron is
misidentified as a photon in  $\PW\to\Pe\PGn$ decays.

\section{Detector, data, and simulated samples}
\label{sec:DataSimSamples}

The data were collected with the CMS detector in 2015. The central feature of
the CMS apparatus is a superconducting solenoid of 6\unit{m} internal diameter,
providing a magnetic field of 3.8\unit{T}. Within the solenoid volume are a
silicon pixel and strip tracker covering the pseudorapidity region $\abs{\eta}<
2.5$, as well as a lead tungstate crystal electromagnetic calorimeter (ECAL),
and a brass and scintillator hadron calorimeter (HCAL), each composed of a
barrel and two endcap sections and covering the range $\abs{\eta}< 3.0$. Forward
calorimeters extend the coverage up to $\abs{\eta}< 5.0$. Muons are measured in
gas-ionization detectors embedded in the iron flux-return yoke outside the
solenoid and cover the range $\abs{\eta}< 2.4$. A more detailed description of the
CMS detector, together with a definition of the coordinate system used and the
relevant kinematic variables, can be found in Ref.~\cite{Chatrchyan:2008zzk}.

The data used in this analysis were selected with a diphoton trigger requiring
a leading photon with $\pt > 30\GeV$ and a subleading photon with $\pt > 18\GeV$.
In order to keep the trigger rate low and to exclude $\cPZ\to\Pe\Pe$ events, a 
combined invariant mass $M_{\gamma\gamma} > 95\GeV$ was also required. 
In addition, the photons were required to pass isolation and cluster shape requirements. 
A sample of $\cPZ\to\Pe\Pe$ events for additional studies was collected with a trigger 
nearly identical to the diphoton trigger, but with an invariant mass requirement 
$M_{\Pe\Pe} > 70\GeV$ and with the additional
requirement that both electromagnetic (EM) objects be matched to a pixel
detector seed (at least two measurements in the pixel detector consistent with
a track from a charged particle).

Monte Carlo simulations of the signal and background processes are used to
validate the performance of the analysis and determine signal efficiencies, as
well as to determine the contributions of some of the smaller backgrounds, as
described in Section 4. The leading-order event generator \MADGRAPH
5.1.3.30~\cite{Alwall:2014hca} is used to simulate the signal samples, which
were generated with either two gluinos or two squarks and up to two additional
partons in the matrix element calculation. The parton showering, hadronization,
multiple-parton interactions, and the underlying event were described by the
\PYTHIA 8~\cite{Sjostrand:2007gs} event generator. The parton distribution
functions are obtained from NNPDF3.0~\cite{Ball:2014uwa}. For the background
processes, the detector response is simulated using
\GEANTfour~\cite{Agostinelli:2002hh}, while the CMS fast
simulation~\cite{Abdullin:2011zz} is used for the signal events.

The signal events were generated using the T5gg and T6gg simplified models and
are characterized by the masses of the particles in the decay chain. For the
gluino (squark) mass we simulate a range of values from 1.0 to 1.8\TeV (1.2 to
2.0\TeV) in steps of 50\GeV. These mass ranges were selected to overlap and
expand upon the mass ranges excluded by previous
searches~\cite{Aad:2015hea,Khachatryan:2015exa,ATLAS:2016aa}. For each gluino (squark) mass, the
\PSGczDo~mass ranges from 100\GeV to 1.9\TeV in 100\GeV increments, with the
requirement that $M_{\PSGczDo} < M_{\PSg}$ ($M_{\PSGczDo} < M_{\PSQ}$). We
assume branching fractions of unity for the decays $\PSg \to
\cPq\cPq\PSGczDo$, $\PSQ \to \cPq\PSGczDo$ and $\PSGczDo \to
\PXXSG\PGg$. For the T6gg model, the gluino mass is set to 10\TeV, and
t-channel production is not considered.

The production cross sections for these processes are calculated as functions
of $M_{\PSg}$ and $M_{\PSQ}$ at next-to-leading-order (NLO) accuracy including
the resummation of soft gluon emission at next-to-leading-logarithmic (NLL)
accuracy~\cite{Kulesza:2009kq, Beenakker:2009ha}, and the uncertainties are
calculated as described in Ref.~\cite{Borschensky:2014cia}.
\section{Event selection}
\label{sec:EventSelect}
Photon, electron, muon, charged hadron, and neutral hadron candidates are
reconstructed with the particle-flow (PF) algorithm~\cite{CMS:2009nxa,
CMS-PAS-PFT-10-001}, which reconstructs particles produced in a collision based
on information from all detector subsystems. Photons are reconstructed from
energy deposits in the ECAL. We require the shape of ECAL clusters to be
consistent with that of an electromagnetic object, and we require that the
energy present in the corresponding region of the HCAL not exceed 5\% of the
ECAL energy, as electromagnetic showers are expected to be contained almost
entirely within the ECAL. In order to ensure that the photons pass the trigger
with high efficiency, all photons are required to satisfy $\et > 40\GeV$.
Because the SUSY signal models used in this analysis produce photons primarily
in the central region of the detector and because the magnitude of the
background increases considerably at high $\abs{\eta}$, we consider only
photons within the barrel fiducial region of the detector ($\abs{\eta} <
1.44$).

To suppress photon candidates originating from quark and gluon jets, photons
are required to be isolated from other reconstructed particles. Separate
requirements are made on the scalar \pt sums of charged hadrons, neutral
hadrons, and electromagnetic objects in a cone of radius $\Delta R \equiv
\sqrt{\smash[b]{(\Delta\eta)^{2} + (\Delta\phi)^{2}}}$ = 0.3 (where $\phi$ is the
azimuthal angle measured in radians) around the photon candidate. Each momentum
sum is corrected for the effect of additional proton-proton collisions in the
event (pileup), and in each case the momentum of the photon candidate itself is
excluded. We further require that the photon candidate have no pixel track
seed, to distinguish the candidate from an electron.

Due to the similarity of the ECAL response to electrons and photons, $\cPZ\to\Pe\Pe$ 
events are used to measure the
photon identification efficiency. The selection of electron candidates is
identical to that of photons, with the exception that the candidate is required
to be matched to a pixel seed consistent with a track, to ensure that the
selection is orthogonal to that of photons. The photon efficiency is measured
via the tag-and-probe method~\cite{Khachatryan:2015iwa} in both data and
simulation. The ratio of the efficiency in data and simulation was measured as
a function of the \pt and $\eta$ of the electron and the $\Delta R$ separation
between the electron and the nearest jet. It is determined that this ratio does
not depend significantly on any measured kinematic variables, and the overall
ratio is computed to be $\epsilon_{\Pe}^{\text{data}} /
\epsilon_{\Pe}^{\text{sim}} = 0.983 \pm 0.012$.

Muon candidates, which are included among the objects counted in the photon
isolation requirement, are reconstructed by performing a global fit that
requires consistent hit patterns in the tracker and the muon
system~\cite{Chatrchyan:2012xi}. We require muons to have $\pt > 30\GeV$ and to
satisfy track quality and isolation requirements. Photons and electrons that 
overlap within $\Delta R < 0.3$ of any muons are rejected, but
otherwise no requirement is made on the number of muons in the event. In addition,
photons must be separated by $\Delta R > 0.3$ from electrons.

Jets are reconstructed from PF candidates using the anti-\kt clustering
algorithm~\cite{Cacciari:2008gp} with a distance parameter of 0.4. The jet
energy and momentum are corrected both for the nonlinear response of the
detector and for the effect of pileup via the procedure described in
Ref.~\cite{1748-0221-6-11-P11002}. Jets are required to have corrected $\pt >
30\GeV$ and to be reconstructed within $\abs{\eta} < 2.4$. In addition, jets
are required to be separated from other objects in the event by $\Delta R >
0.4$.

For the purpose of defining the various control regions used in the analysis,
we apply an additional set of selection criteria. Misidentified photons are
defined as those photon candidates passing the photon selection but failing
either the shape requirement for the ECAL clusters or the charged-hadron
isolation requirement, but not both. In order to ensure that misidentified
photons do not differ too much from our photon selection, upper limits are
applied to both the charged-hadron isolation and cluster shape requirements.

Events are then sorted into one of four mutually exclusive categories depending
on the selection of their highest-\pt electromagnetic objects: $\gamma\gamma$,
ee, two misidentified (``fake'') photons (ff), and $\Pe\gamma$. Due to the
trigger requirements described in Section 2, the invariant mass of the two
electromagnetic objects is required to be greater than 105\GeV. The size
of the data sample limits any improvements in the sensitivity of the analysis
from categorizing events by jet multiplicity. Therefore, no requirements are
made on the number of jets in the event.

The signal region is defined by the events in the $\gamma\gamma$ category with
$\ETmiss \geq 100\GeV$ and is split into four bins: $100 \leq \ETmiss <
110\GeV$, $110 \leq \ETmiss < 120\GeV$, $120 \leq \ETmiss < 140\GeV$, and
$\ETmiss \geq 140\GeV$. The bins are chosen in such a way that there is a
sufficient amount of data in each bin in the ee and ff control samples used for
background estimation. The bin with $\ETmiss < 100\GeV$ is used as a control
region.
\section{Estimation of backgrounds}
\label{sec:EstBkg}

The dominant background for this analysis comes from multijet production from
quantum chromodynamics (QCD) processes without intrinsic \ETmiss, where the
high-\ETmiss signature is mimicked by the mismeasurement of the hadronic
activity in the event. A subdominant contribution comes from electroweak (EWK)
processes that include intrinsic \ETmiss from neutrino production.

\begin{figure}[htbp]
    \centering
    \includegraphics[width=\cmsFigWidth]{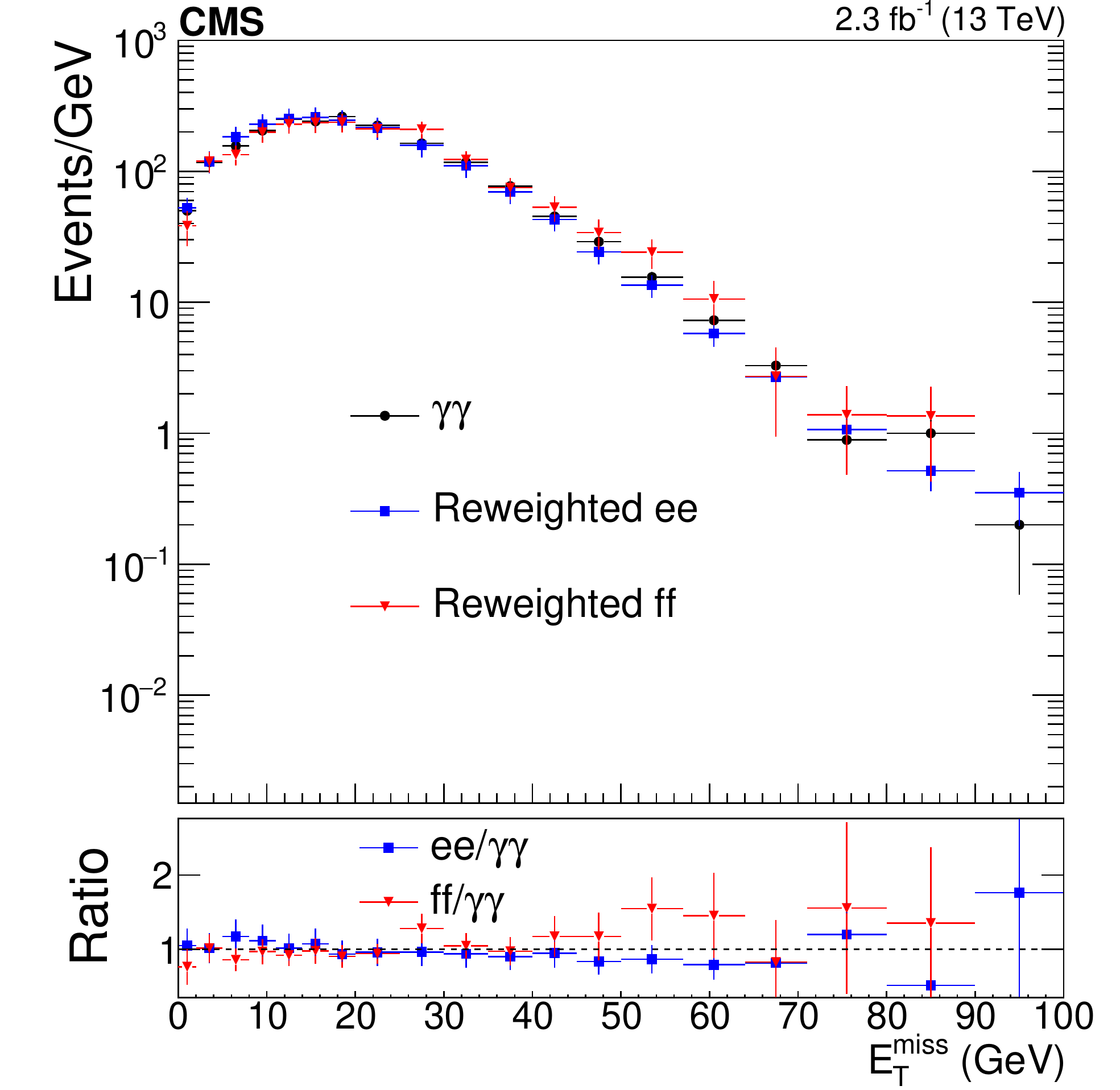}
    \caption{The \ETmiss distributions of the candidate $\gamma\gamma$,
        reweighted ee, and reweighted ff samples in the $\ETmiss < 100\GeV$
        sideband.}
    \label{fig:reweightEffect}
\end{figure}

The contribution from the QCD background is modeled in a fully data-driven way
from the ee and ff control samples. Both of these control samples are dominated
by processes without intrinsic \ETmiss and can therefore be used to model the
\ETmiss in the QCD background. These control samples differ in hadronic
activity from the candidate $\gamma\gamma$ sample due to different event
topologies. In particular, the ee control sample has a large contribution from
$\cPZ \to \Pe\Pe$ events, where the
electromagnetic objects come from one parent particle. In contrast, the ff
control samples are primarily multijet events where the two electromagnetic
objects are produced independently. To account for this difference, the di-EM
\pt variable, defined as the magnitude of the vector sum of the transverse
momenta of the two electromagnetic objects, is used to model the hadronic
recoil in the event. Events in the ee and ff control samples are reweighted by
the di-EM \pt distribution of the $\gamma\gamma$ events to correct for any
differences in hadronic recoil.
The \ETmiss distributions of these di-EM \pt reweighted control samples are
then normalized to that of the $\gamma\gamma$ sample in the region $\ETmiss <
50\GeV$ and used to predict the contribution of QCD processes to the
high-\ETmiss signal region. A comparison of the reweighted \ETmiss
distributions to the distribution of $\gamma\gamma$ events is shown in
Fig.~\ref{fig:reweightEffect} in the sideband of the search region ($\ETmiss <
100\GeV$). There is an agreement within statistical uncertainties between the
$\gamma\gamma$ and each of the reweighted distributions.

Similarly, we consider differences in the \ETmiss distribution due to the
number of jets in the event. A direct comparison of the candidate sample and
the two control samples shows little dependence on the jet multiplicity
$N_{\text{jets}}$ at low \ETmiss, so we take the difference as a systematic
uncertainty in the prediction, as described in Section 5.

In addition, there is a small contribution in the QCD control samples from
comparatively rare processes with intrinsic \ETmiss, including
\cPqt\cPaqt~events and $\cPZ \to \Pgn\Pagn$ + jets events. Due to their
small cross sections, these processes are estimated with simulation, and their
contributions are subtracted from the ee and ff control samples for the final
prediction.

The primary estimate of the QCD contribution comes from the reweighted ee
distribution. The reweighted ff control sample serves as a cross-check, and the
difference between them is taken as a symmetric systematic uncertainty on the
prediction. Due to the limited number of ff events with $\ETmiss > 100\GeV$, a
looser misidentification definition is used. In the looser definition,
misidentified photons are not required to pass any photon isolation or
neutral-hadron isolation cuts, and the upper limits on charged-hadron isolation
and the shape requirement for the ECAL clusters are loosened further. The
looser ff sample is used to obtain the shape of the ff distribution in the
$\ETmiss > 100\GeV$ signal region, while the normalization comes from the
tighter, more photon-like misidentification definition.

As an additional cross-check on this background estimation method, the ratio
of the candidate $\gamma\gamma$ distribution to the unweighted ff
distribution as a function of \ETmiss is fit with different functional forms.
The predicted number of QCD background events in each \ETmiss bin is then 
given by the function multiplied by the number of ff events seen in that bin.
The primary prediction from the ee sample is consistent with the prediction
from this cross-check within the fit uncertainties, and we conclude that the
predictions from these two methods are compatible.

The electroweak background comes from $\PW\gamma$ events where the W decays to an
electron and a neutrino, and the electron is misidentified as a photon. We
estimate this misidentification rate by comparing the Z-boson mass peak in the
ee invariant mass spectrum to the peak in the $\Pe\gamma$ spectrum. The data are
modeled using an extended likelihood fit to the mass spectrum for the signal
plus background hypothesis. The misidentification rate $f_{\Pe
\to \gamma}$ is then computed from the signal events as $f_{\Pe
\to \gamma} = N_{\Pe\gamma} / (2N_{\Pe\Pe} +
N_{\Pe\gamma}) = (2.13 \pm 0.21)\%$. This rate is used to compute a
scaling factor $f_{\Pe \to \gamma} / (1 - f_{\Pe
\to \gamma})$, which is then applied to the sample of $\Pe\gamma$ events
with $\ETmiss > 100\GeV$ to obtain an estimate of the electroweak background in
the signal region.
\section{Sources of systematic uncertainty}
\label{sec:SystUncert}

We evaluate systematic uncertainties from each of the background predictions,
the signal efficiency, and the integrated luminosity. For each source of
uncertainty, we give the uncertainty value and describe the method used for its
estimation.

The largest systematic uncertainties come from the QCD background estimation
method. We consider three sources of systematic uncertainty from the QCD
background estimate: the di-EM \pt reweighting, the jet multiplicity
dependence, and the \ETmiss shape difference between the $\Pe\Pe$ and ff control
samples. The magnitudes of these uncertainties for each of the \ETmiss bins in
the signal region are shown in Table~\ref{tab:SystUncerts}.

\begin{table*}[ht]
  \topcaption{Systematic uncertainties from the QCD background estimation.}
  \centering
  \begin{tabular}{ rcccc }
 \multicolumn{1}{c}{$\ETmiss$ bin ($\GeVns{}$)} & Di-EM $\pt$ & Jet multiplicity & Shape difference & Statistical uncertainty \\ [0.5ex]
 & reweighting & reweighting & between ee and ff & of ee sample \\ [0.5ex]
 \hline
 $100 \leq \ETmiss < 110$ & 15\% & 34\% & 18\% & 31\% \\
 $110 \leq \ETmiss < 120$ & 17\%& 15\% & 12\% & 33\% \\
 $120 \leq \ETmiss < 140$ & 33\%& 29\% & 14\% & 42\% \\
 $\ETmiss \geq 140$ & 39\%& 20\% & 150\% & 71\% \\
\end{tabular}
\label{tab:SystUncerts}
\end{table*}

The uncertainty from di-EM \pt reweighting is estimated from the distributions
of the di-EM \pt ratio in simulated pseudo-experiments, allowing the ratio to
vary bin by bin according to a Gaussian distribution with a standard deviation
computed from the statistical uncertainty of unweighted events in the bin. The
\ETmiss distribution of the ee control sample is then reweighted by each of
these distributions, and the standard deviation is determined for the
prediction. The magnitude of this uncertainty ranges from 15\% to 39\%.

The effect of the difference in the \ETmiss distribution as a function of the
jet multiplicity is determined directly by taking the difference between the ee
estimate with di-EM \pt and $N_{\text{jets}}$ reweighting and with di-EM \pt
reweighting alone. The resulting systematic uncertainty ranges from 15\% to
34\% in the four signal \ETmiss bins. The shape uncertainty of the ee control
sample is determined by fitting the high-\ETmiss tails of the ee and ff samples
to the empirical three-parameter function $\ddinline{N}{\ETmiss} =
(\ETmiss)^{p_{0}} \re^{p_{1}(\ETmiss)^{p_{2}}}$. The systematic uncertainty in
the shape is symmetric and taken to be the fractional difference in each
\ETmiss bin between these fitted functions. This yields a systematic effect
between 12\% and 18\% in the lower three \ETmiss signal bins, and a systematic
effect of 150\% in the final bin that covers \ETmiss above 140\GeV.

The main source of uncertainty in the electroweak background estimate comes
from the uncertainty in the extended likelihood fit used to calculate the
misidentification rate. This is computed by shifting the rate up and down by
its uncertainty and scaling the \ETmiss distribution of the $\Pe\gamma$ control
sample by the altered rates. The difference between the estimates from the two
shifted misidentification rates gives the systematic uncertainty in the rate of
electroweak events. Because this represents an uncertainty in the overall
normalization, it is constant across \ETmiss bins. The uncertainty is a
constant 19\% across the \ETmiss bins.

The signal efficiency uncertainties are related to the statistical uncertainty
from the finite size of the T5gg and T6gg signal samples (0--16\%), knowledge
of the jet energy scale (0--23\% depending on the \PSg--\PSGczDo mass
difference), parton distribution function uncertainties (13--22\% depending on
the signal point), and photon identification and reconstruction efficiencies
(2\%). The uncertainty related to the integrated luminosity of the data
sample is 2.7\%~\cite{CMS:2016eto}.
\section{Results}
\label{sec:Results}

\begin{figure}[!htbp]
    \centering
    \includegraphics[width=\cmsFigWidth]{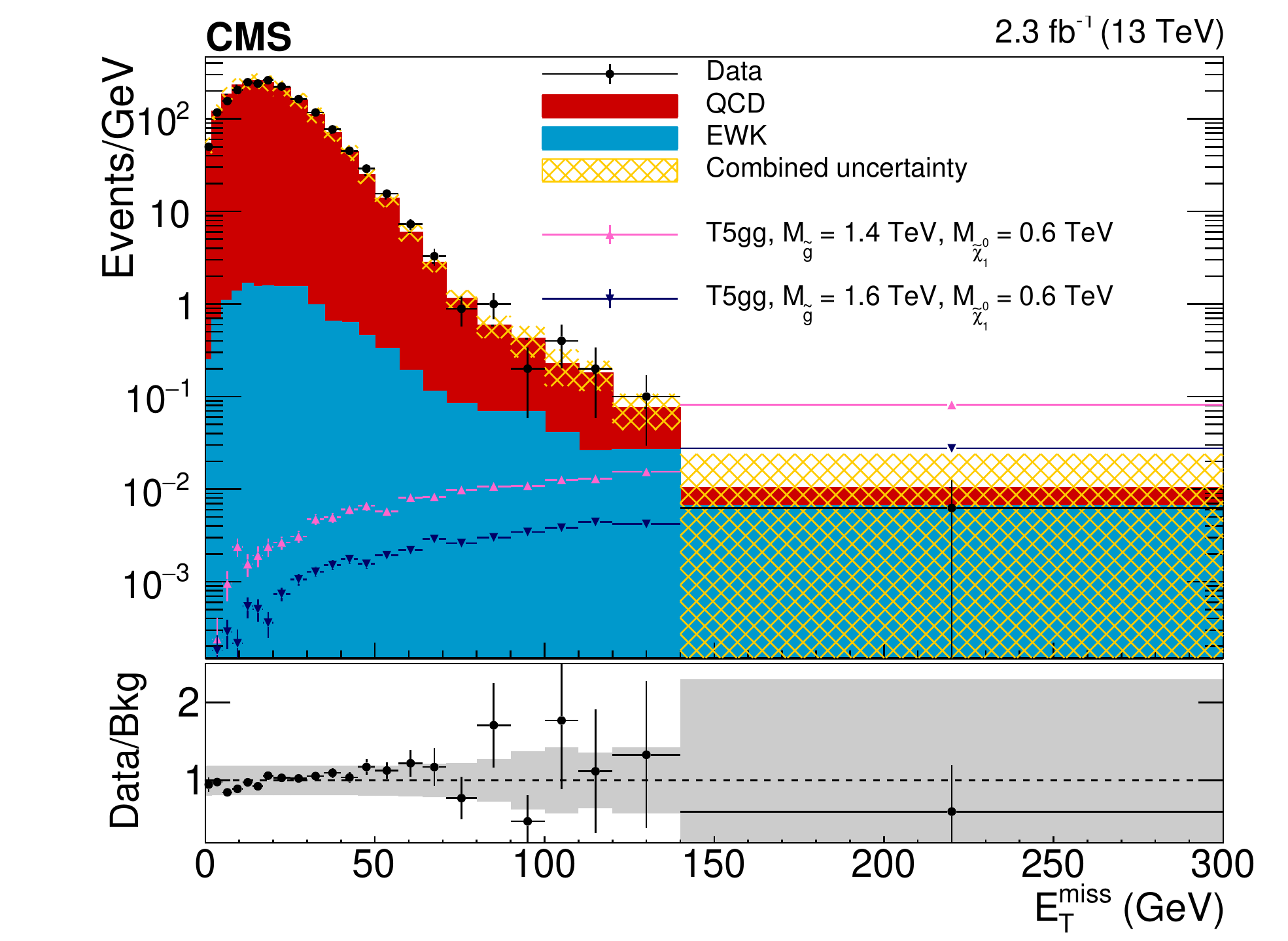}
    \caption{Measured \ETmiss distribution in comparison with the background
        prediction. The four bins with $\ETmiss \geq 100\GeV$ constitute the
        signal region, and the $\ETmiss < 100 \GeV$ bins serve mainly to
        normalize the background. The systematic uncertainty on the background
        prediction and the ratio of the data to the prediction are also shown.
        The last bin includes all events with $\ETmiss \geq 140\GeV$, but for
        normalization by bin width, the bin is taken to be from $140 \leq
        \ETmiss < 300\GeV$. The distributions for two signal model points are
        overlaid for comparison.}

    \label{fig:metDistribution}
\end{figure}

The measured \ETmiss distribution and corresponding background predictions are
shown in Fig.~\ref{fig:metDistribution}. The expected number of events from the
QCD and EWK backgrounds, as well as the total number of expected and observed
events, are shown in Table~\ref{tab:ExpObs} for each bin in the signal region.
We observe 9 events total in the signal region, compared to an expected
background of $7.2\pm2.5$ events. The number of events in the signal region
agrees with the background estimate within the uncertainties.

\begin{table*}[htb]
    \topcaption{Numbers of expected and observed events in the signal region. The
        last row shows the total number of expected and observed events in the
        inclusive bin $\ETmiss \geq 100\GeV$. The expected numbers of 
        events for two T5gg mass points are also shown. For Signal A, 
        $M_{\PSg} = 1400\GeV$ and $M_{\PSGczDo} = 600\GeV$. For Signal B,
        $M_{\PSg} = 1600\GeV$ and $M_{\PSGczDo} = 600\GeV$.
        The uncertainties include all of
        the systematic uncertainties described in
        Section~\ref{sec:SystUncert}. }
    \centering
    \begin{tabular}{ rcccccc }
        \multicolumn{1}{c}{$\ETmiss$ bin ($\GeVns{}$) }& QCD & EWK & Total background & Signal A & Signal B & Observed\\ [0.5ex]
        \hline
        $100 \leq \ETmiss < 110$ & 1.9$\pm$1.0 & 0.4$\pm$0.1 & 2.3$\pm$1.0 & 0.12$\pm$0.01 & 0.04$\pm$0.01 & 4 \\
        $110 \leq \ETmiss < 120$ & 1.5$\pm$0.6 & 0.3$\pm$0.1 & 1.8$\pm$0.6 & 0.13$\pm$0.02 & 0.04$\pm$0.01 & 2 \\
        $120 \leq \ETmiss < 140$ & 1.0$\pm$0.6 & 0.5$\pm$0.2 & 1.5$\pm$0.6 & 0.31$\pm$0.04 & 0.08$\pm$0.01 & 2 \\
        $\ETmiss \geq 140$ & 0.6$\pm$2.2 & 1.0$\pm$0.3 & 1.6$\pm$2.2 & 13.0$\pm$0.7 & 4.4$\pm$0.2 & 1 \\
        $\ETmiss \geq 100$ & 5.0$\pm$2.5 & 2.2$\pm$0.3 & 7.2$\pm$2.5 & 13.6$\pm$0.7 & 4.6$\pm$0.2 & 9 \\
    \end{tabular}
    \label{tab:ExpObs}
\end{table*}

We determine 95\% confidence level (CL) upper limits on gluino pair and squark
pair production cross sections using the modified frequentist CL$_{\mathrm{s}}$
method~\cite{Junk:1999kv, Read:2002hq} based on a log-likelihood test statistic
that compares the likelihood of the SM-only hypothesis to the likelihood of the
presence of a signal in addition to the SM contributions. The likelihood
function is constructed from the background and signal \ETmiss distributions
across the four bins described in Section~3. The systematic uncertainties
described in Section~\ref{sec:SystUncert} are included in the test statistic as
nuisance parameters, with log-normal probability distributions.

\begin{figure}[htbp]
    \centering
    \includegraphics[width=0.48\textwidth]{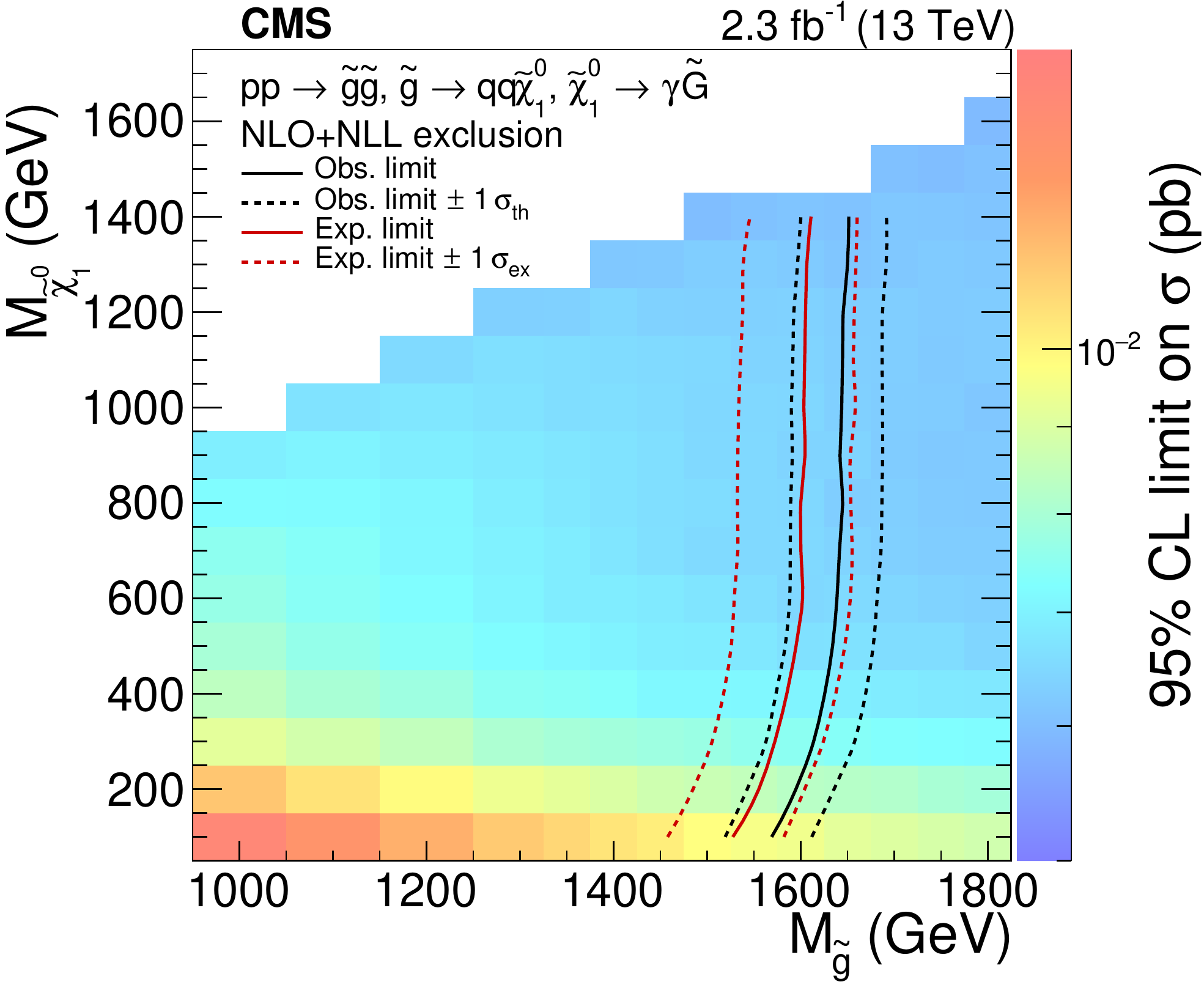}
    \includegraphics[width=0.48\textwidth]{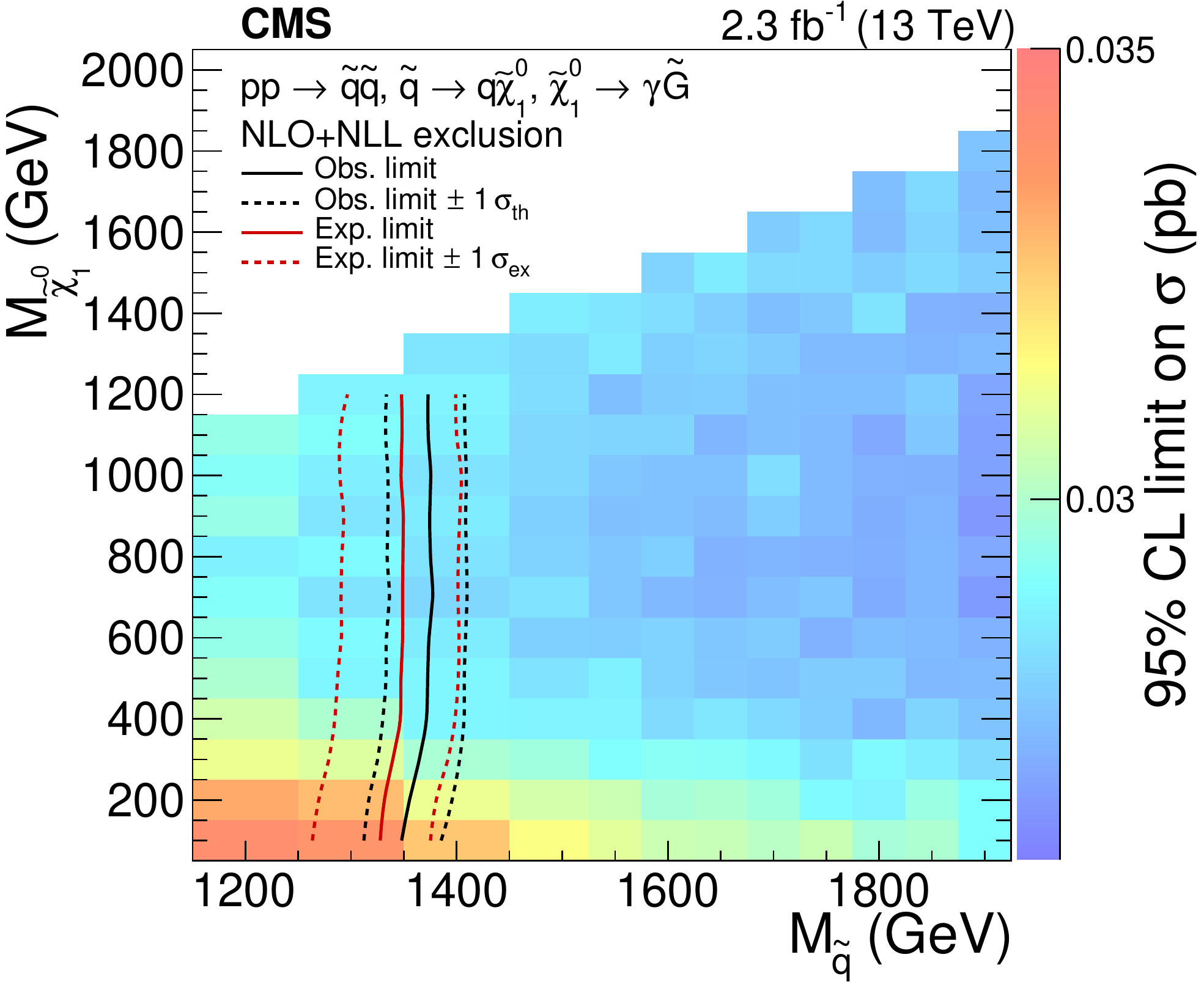}
    \caption{The 95\% CL upper limits on the gluino (\cmsLeft) and squark (\cmsRight)
        pair production cross sections as a function of neutralino versus
        gluino (squark) mass. The contours show the observed and median
        expected exclusions assuming the NLO+NLL cross sections, with their one
        standard deviation uncertainties. The limit curves terminate at the
        centers of the bins used to sample the cross section.}
    \label{fig:limit}
\end{figure}

In Fig.~\ref{fig:limit} we present 95\% CL upper limits on the cross section as
a function of the mass pair values for the two models considered in this
analysis, $M_{\PSGczDo}$ versus $M_{\PSg}$ and $M_{\PSGczDo}$ versus $M_{\PSQ}$
for gluino pair and squark pair production, respectively. From the NLO+NLL
predicted cross sections and their uncertainties we derive contours
representing lower limits in the SUSY mass plane. We also show expected limit
contours based on the expected experimental cross section limits and their
uncertainties. For typical values of the neutralino mass, we expect to exclude
gluino masses up to 1.60\TeV and squark masses up to 1.35\TeV, and we observe
exclusions of 1.65 and 1.37\TeV respectively. The excluded mass ranges for 
gluino pair production have been improved by approximately 300\GeV with 
respect to previous searches performed at $\sqrt{s} = 8\TeV$~
\cite{Aad:2015hea,Khachatryan:2015exa}. The observed 
exclusions are consistent with the results of the ATLAS
analysis performed at $\sqrt{s} = 13\TeV$~\cite{ATLAS:2016aa}.

\section{Summary}
\label{sec:Summary}

A search is performed for supersymmetry with general gauge mediation in
proton-proton collisions yielding events with two photons and large missing
transverse energy. The data were collected at a center-of-mass energy of 13\TeV
with the CMS detector in 2015, and correspond to an integrated luminosity of
2.3\fbinv.

The data are interpreted in the context of two simplified SUSY models
with gauge-mediated supersymmetry breaking, one assuming
gluino pair production and the second assuming squark pair production.
In both models, the branching fraction of the NLSP neutralino to decay to a
gravitino and a photon is assumed to be unity.
Using background estimation methods based on control samples in data, limits
are determined on the gluino and squark pair production cross sections, and
those limits are used together with NLO+NLL cross section calculations to
constrain the masses of gluinos, squarks, and neutralinos.
Gluino masses below 1.65\TeV and
squark masses below 1.37\TeV are excluded at a 95\% confidence level.
This represents an improvement of approximately 300\GeV with respect to previous
analyses performed at a center-of-mass energy of 8\TeV~\cite{Aad:2015hea,Khachatryan:2015exa}
and is consistent with the results of the ATLAS analysis performed at a center-of-mass energy 
 of 13\TeV ~\cite{ATLAS:2016aa}.

\begin{acknowledgments}
We congratulate our colleagues in the CERN accelerator departments for the excellent performance of the LHC and thank the technical and administrative staffs at CERN and at other CMS institutes for their contributions to the success of the CMS effort. In addition, we gratefully acknowledge the computing centres and personnel of the Worldwide LHC Computing Grid for delivering so effectively the computing infrastructure essential to our analyses. Finally, we acknowledge the enduring support for the construction and operation of the LHC and the CMS detector provided by the following funding agencies: BMWFW and FWF (Austria); FNRS and FWO (Belgium); CNPq, CAPES, FAPERJ, and FAPESP (Brazil); MES (Bulgaria); CERN; CAS, MoST, and NSFC (China); COLCIENCIAS (Colombia); MSES and CSF (Croatia); RPF (Cyprus); SENESCYT (Ecuador); MoER, ERC IUT, and ERDF (Estonia); Academy of Finland, MEC, and HIP (Finland); CEA and CNRS/IN2P3 (France); BMBF, DFG, and HGF (Germany); GSRT (Greece); OTKA and NIH (Hungary); DAE and DST (India); IPM (Iran); SFI (Ireland); INFN (Italy); MSIP and NRF (Republic of Korea); LAS (Lithuania); MOE and UM (Malaysia); BUAP, CINVESTAV, CONACYT, LNS, SEP, and UASLP-FAI (Mexico); MBIE (New Zealand); PAEC (Pakistan); MSHE and NSC (Poland); FCT (Portugal); JINR (Dubna); MON, RosAtom, RAS, and RFBR (Russia); MESTD (Serbia); SEIDI and CPAN (Spain); Swiss Funding Agencies (Switzerland); MST (Taipei); ThEPCenter, IPST, STAR, and NSTDA (Thailand); TUBITAK and TAEK (Turkey); NASU and SFFR (Ukraine); STFC (United Kingdom); DOE and NSF (USA).

\hyphenation{Rachada-pisek} Individuals have received support from the Marie-Curie programme and the European Research Council and EPLANET (European Union); the Leventis Foundation; the A. P. Sloan Foundation; the Alexander von Humboldt Foundation; the Belgian Federal Science Policy Office; the Fonds pour la Formation \`a la Recherche dans l'Industrie et dans l'Agriculture (FRIA-Belgium); the Agentschap voor Innovatie door Wetenschap en Technologie (IWT-Belgium); the Ministry of Education, Youth and Sports (MEYS) of the Czech Republic; the Council of Science and Industrial Research, India; the HOMING PLUS programme of the Foundation for Polish Science, cofinanced from European Union, Regional Development Fund, the Mobility Plus programme of the Ministry of Science and Higher Education, the National Science Center (Poland), contracts Harmonia 2014/14/M/ST2/00428, Opus 2013/11/B/ST2/04202, 2014/13/B/ST2/02543 and 2014/15/B/ST2/03998, Sonata-bis 2012/07/E/ST2/01406; the Thalis and Aristeia programmes cofinanced by EU-ESF and the Greek NSRF; the National Priorities Research Program by Qatar National Research Fund; the Programa Clar\'in-COFUND del Principado de Asturias; the Rachadapisek Sompot Fund for Postdoctoral Fellowship, Chulalongkorn University and the Chulalongkorn Academic into Its 2nd Century Project Advancement Project (Thailand); and the Welch Foundation, contract C-1845.

\end{acknowledgments}
\bibliography{auto_generated}

\ifthenelse{\boolean{cms@external}}{}{
\clearpage
\appendix
\numberwithin{table}{section}
\numberwithin{figure}{section}
\section{Supplemental information\label{app:suppMat}}
\subsection{Contributions to control samples}
As described in \ifthenelse{\boolean{cms@external}}{Section~4 of the main document}{Section~\ref{sec:EstBkg}}, the small contribution to the $\Pe\Pe$ control sample
from \cPqt\cPaqt~events and the contribution to the ff sample from
$\cPZ \to \Pgn\Pagn$ + jets events are
estimated with simulation. The size of the contributions are listed in
Table~\ref{tab:BkgContributions}.

 \begin{table}[ht]
  \topcaption{Percent contributions from background samples to the $\Pe\Pe$ and ff control samples.}
  \centering
  \begin{tabular}{ lcc }
$\ETmiss$ &  $<$100\GeV & $\geq$100\GeV \\
 \cline{2-3}
 $\cPqt\cPaqt$ events & 0.17\% &  24.3\% \\
 $\cPZ \to\Pgn\Pagn$ + jets events  & 0.03\% & 5.0\% \\
 \hline
\end{tabular}
\label{tab:BkgContributions}
\end{table}

\subsection{Reweighting distributions}
Figure~\ref{fig:diEmPtRatio} shows the di-EM \pt distributions of the
$\gamma\gamma$ candidate sample and ee and ff control samples, as well
as the ratios used for reweighting.

\begin{figure}[tbh]
    \centering
    \includegraphics[width=\cmsFigWidth]{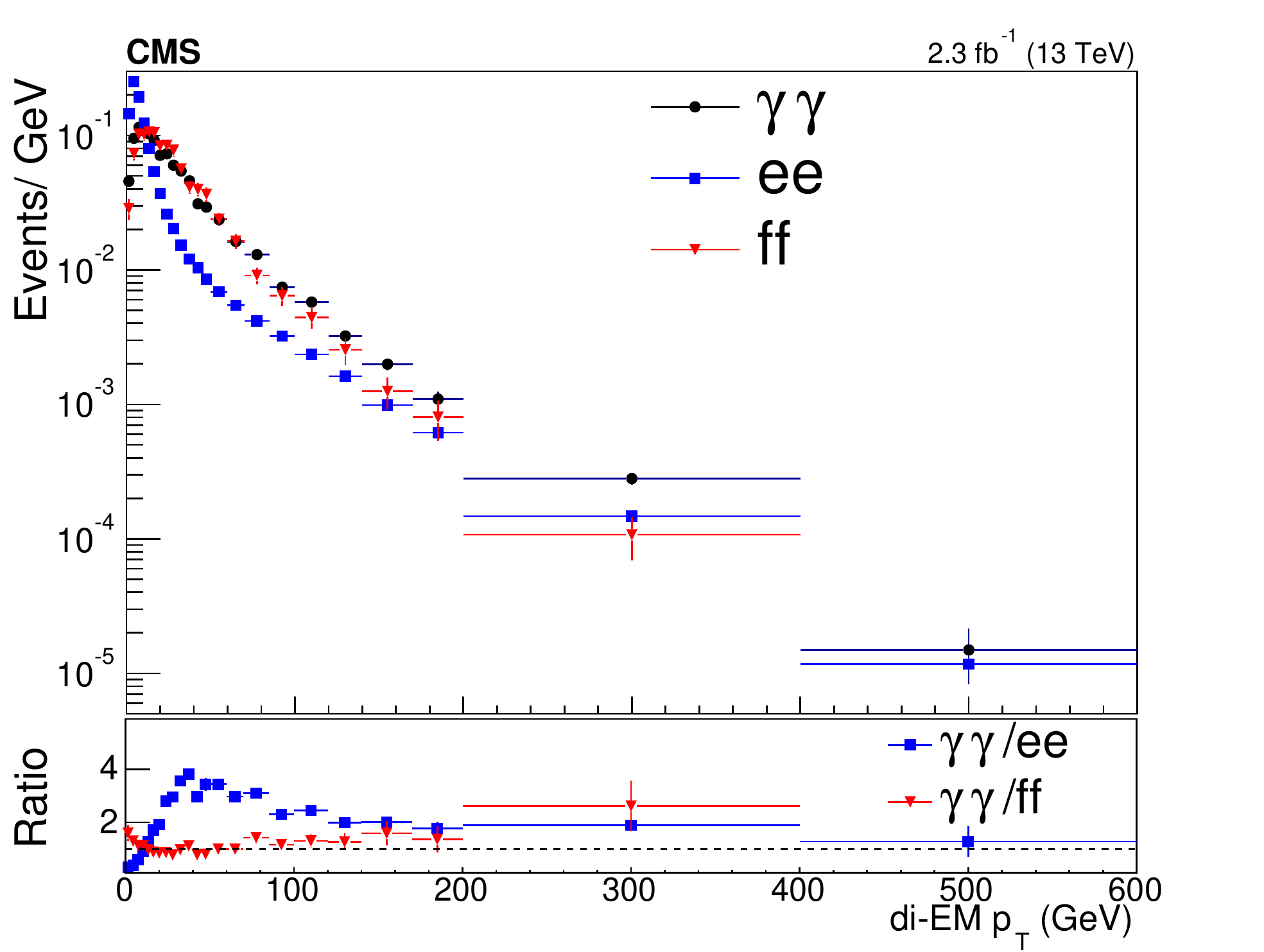}
    \caption{Di-EM \pt distribution of the $\gamma\gamma$ candidate sample and
         $\Pe\Pe$ and ff control samples. The ratios of the candidate sample to each
         of the control samples are shown in the bottom pane. These ratios serve as the
         reweighting factors for the events.}
     \label{fig:diEmPtRatio}
\end{figure}

}

\cleardoublepage \section{The CMS Collaboration \label{app:collab}}\begin{sloppypar}\hyphenpenalty=5000\widowpenalty=500\clubpenalty=5000\textbf{Yerevan Physics Institute,  Yerevan,  Armenia}\\*[0pt]
V.~Khachatryan, A.M.~Sirunyan, A.~Tumasyan
\vskip\cmsinstskip
\textbf{Institut f\"{u}r Hochenergiephysik,  Wien,  Austria}\\*[0pt]
W.~Adam, E.~Asilar, T.~Bergauer, J.~Brandstetter, E.~Brondolin, M.~Dragicevic, J.~Er\"{o}, M.~Flechl, M.~Friedl, R.~Fr\"{u}hwirth\cmsAuthorMark{1}, V.M.~Ghete, C.~Hartl, N.~H\"{o}rmann, J.~Hrubec, M.~Jeitler\cmsAuthorMark{1}, A.~K\"{o}nig, I.~Kr\"{a}tschmer, D.~Liko, T.~Matsushita, I.~Mikulec, D.~Rabady, N.~Rad, B.~Rahbaran, H.~Rohringer, J.~Schieck\cmsAuthorMark{1}, J.~Strauss, W.~Waltenberger, C.-E.~Wulz\cmsAuthorMark{1}
% ...
\vskip\cmsinstskip
% ...
1:~~Also at Vienna University of Technology, Vienna, Austria\\
% ...

\end{sloppypar}
\end{document}